\documentclass{svjour3}                     
\smartqed  
\usepackage{amssymb}
\usepackage{graphicx}
\usepackage{natbib}
\bibpunct{(}{)}{;}{a}{}{,}

\newcommand{\aeta}[3]{  #1, { A\&A}, {  #2}, #3}

\newcommand{\araa}[3]{  #1, { ARA\&A}, {  #2}, #3}
\newcommand{\aj}[3]{  #1, { AJ}, {  #2}, #3}
\newcommand{\aspj}[3]{  #1, { ApJ}, {  #2}, #3}

\newcommand{\aspjs}[3]{  #1, { ApJS}, {  #2}, #3}

\newcommand{\mnras}[3]{   #1, { MNRAS}, {  #2}, #3}

\newcommand{\rvmp}[3]{   #1, { Rev. Mod. Phys.}, {  #2},  #3}

\newcommand{\SpaceS}[3]{#1, { Space Sci. Rev.}, {  #2}, #3}

 \usepackage{mathptmx}      
\newcommand{\ion}[2]{#1\,{\sc{#2}}}
 \journalname{SSRv}
\begin{document}

\title{Clusters of galaxies: Setting the stage
}


\author{A. Diaferio \and
        S. Schindler \and 
	K. Dolag 
}

\authorrunning{Diaferio et al.} 

\institute{A. Diaferio \at
              Dipartimento di Fisica Generale ``Amedeo Avogadro'', Universit\`a degli Studi di Torino,
                  Via P. Giuria 1, I-10125, Torino, Italy \\
Istituto Nazionale di Fisica Nucleare (INFN), Sezione di Torino, Via P. Giuria 1, I-10125, Torino, Italy \\
              \email{diaferio@ph.unito.it}           
           \and
           S. Schindler \at
             Institut f\"ur Astro- und Teilchenphysik der Leopold-Franzens Universit\"at Innsbruck,
                  Technikerstrasse 25/8, A-6020, Innsbruck, Austria \\
	\and
	K. Dolag \at 
Max-Planck-Institut f\"ur Astrophysik,
                  Karl-Schwarzschildstr. 1, P.O. Box 1317, D-85741, Garching b. M\"unchen, Germany
}

\date{Received: 21 September 2007 / Accepted: 28 September 2007}

\maketitle

\begin{abstract}
Clusters of galaxies are self-gravitating systems of mass $\sim 10^{14}-10^{15} h^{-1}$ M$_\odot$ and size
$\sim 1-3 h^{-1}$ Mpc. Their mass budget consists of dark matter ($\sim 80\%$, on average),
hot diffuse intracluster plasma ($\lesssim 20\%$) and a small fraction of stars, dust, and cold gas, mostly locked
in galaxies. In most clusters, scaling relations between their
properties, like mass, galaxy velocity dispersion, X-ray luminosity and temperature, testify that the cluster
components are in approximate
dynamical equilibrium within the cluster gravitational potential well.
However, spatially inhomogeneous thermal and non-thermal emission of the
intracluster medium (ICM),
observed in some clusters in the X-ray and radio bands,
and the kinematic and morphological segregation of galaxies
are a signature of non-gravitational processes, ongoing cluster merging and interactions.
Both the fraction of clusters with these features, and
the correlation between the dynamical and morphological properties of
irregular clusters and the surrounding large-scale structure increase with redshift.

In the current bottom-up scenario for the formation of cosmic structure, where
tiny fluctuations of the otherwise
homogeneous primordial density field are amplified by gravity, clusters are
the most massive nodes of the filamentary large-scale structure of the
cosmic web and form by anisotropic and episodic accretion of mass,
in agreement with most of the observational evidence.
In this model of the universe dominated by cold dark matter, at the present time most baryons are
expected to be in a diffuse component rather than in stars and galaxies; moreover, $\sim 50\%$ of this diffuse component
has temperature $\sim 0.01-1$ keV and permeates the filamentary
distribution of the dark matter. The temperature of this Warm-Hot Intergalactic Medium (WHIM) increases
with the local density and its search
in the outer regions of clusters and lower density regions has been the quest of much recent observational effort.

Over the last thirty years, an impressive coherent picture of the
formation and evolution of cosmic structures has emerged from the intense interplay between observations, theory
and numerical experiments. Future efforts will continue
to test whether this picture keeps being valid, needs corrections or
suffers dramatic failures in its predictive power.

\end{abstract}

\section{Preamble}

The present chapter provides the general framework of the physics of
clusters of galaxies: it outlines how clusters connect to several astrophysical
issues, from cosmology to the formation of galaxies and stars. 
Some of these topics are discussed in the chapters of this volume;
we refer to these chapters when appropriate. 
When the topic we mention here is not dealt with elsewhere in this volume, 
we refer to some of the most recent reviews.
   
\section{The observational framework }
\label{sec:obs}

\subsection{Cluster components}
\label{subsec:clustcomp}

\subsubsection{Mass content}

The discovery of clusters as concentrations of galaxies on the sky 
dates back to 1784, when Charles Messier mentions 
the Virgo cluster in his {\it Connaissance des Temps} \citep[see the fascinating historical review by][]{biviano00}.  
The first systematic optical surveys of clusters appeared almost
two centuries later \citep{abell58, zwicky68}.

\begin{figure}    
\begin{center}
\hbox{
\includegraphics[width=6.cm]{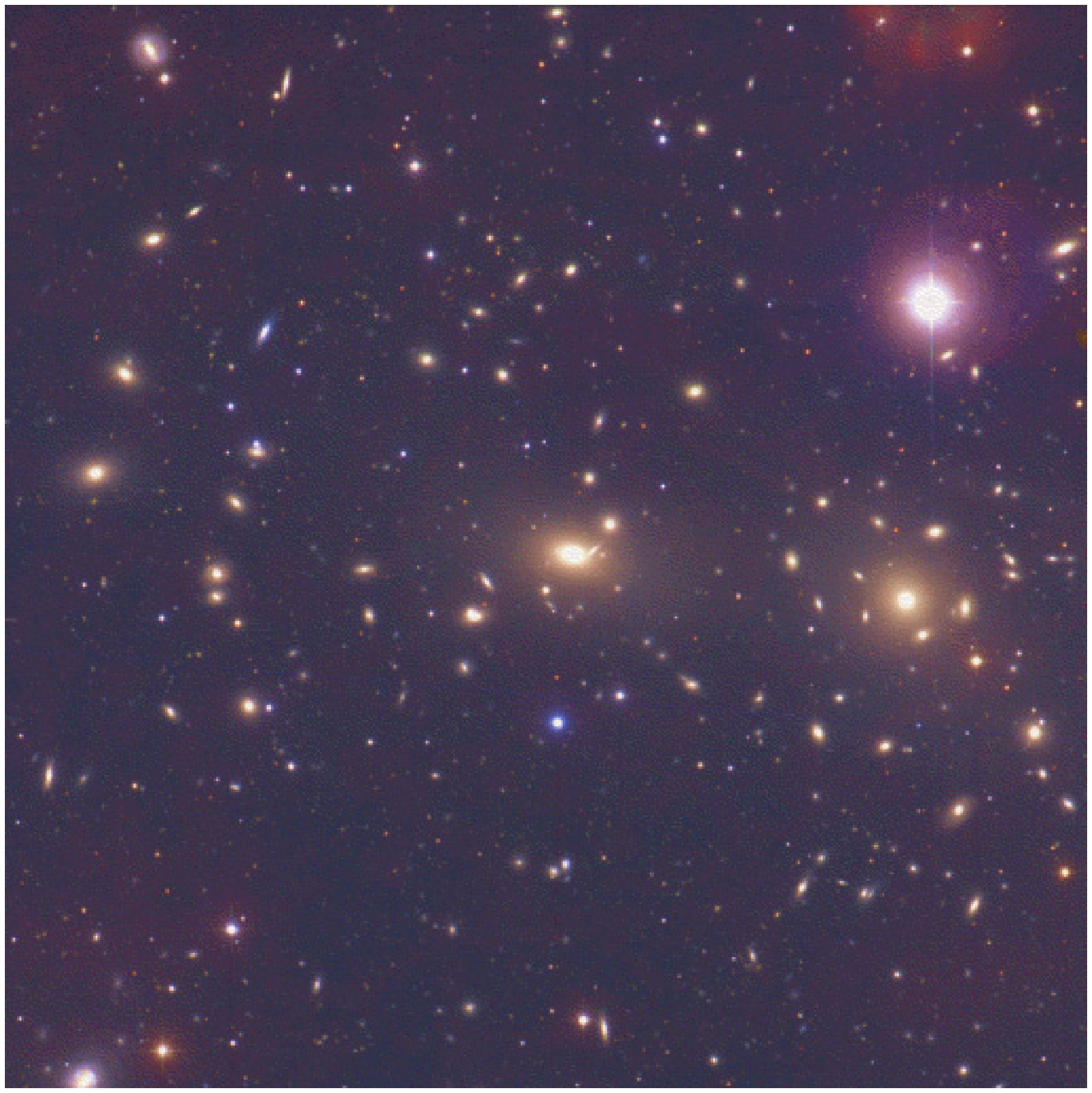}
\includegraphics[width=6.cm]{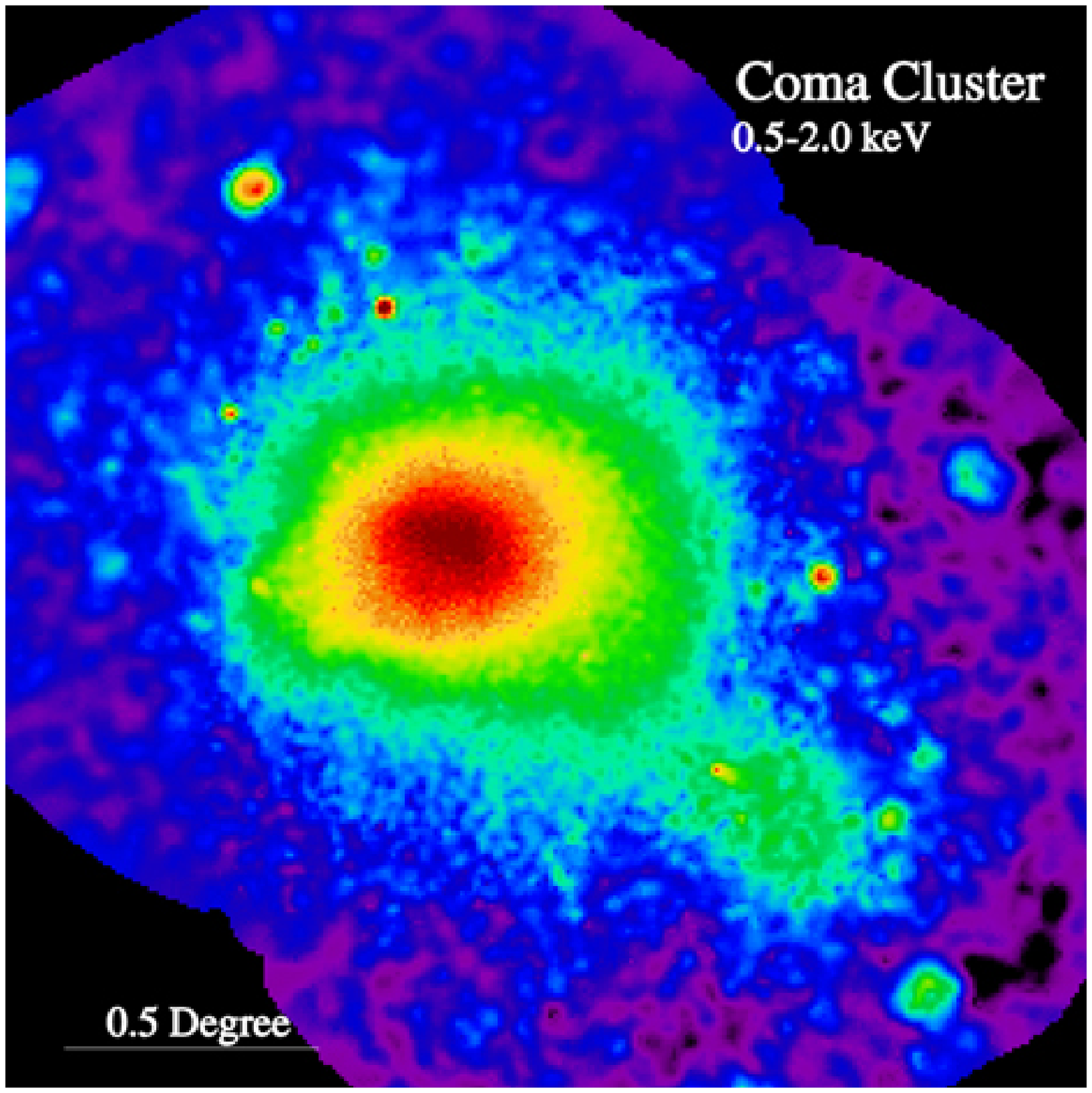}
}
\caption{Left panel: Optical image of the central part of the Coma cluster ($23\times 23$ arcmin)
centered on the galaxy NGC4889; the bright galaxy on the right is NGC4874. Right panel: 
X-ray image (ROSAT) of Coma ($164\times 150$ arcmin) 
(from http://chandra.harvard.edu/photo/2002/0150/more.html).  }
\label{fig:Coma}
\end{center}
\end{figure}

In 1930's, Zwicky, by assuming that the cluster in the Coma constellation, 
containing hundreds of bright galaxies (Fig.~\ref{fig:Coma}),\footnote{See \cite{biviano98} for
a historical perspective on the Coma cluster.}  
is in virial equilibrium, found that the mass required 
to bind the system gravitationally should be roughly hundred times larger
than the sum of the masses of the individual galaxies \citep{zwicky33, zwicky37}. In fact, if the cluster is
an isolated, spherically symmetric system in dynamical equilibrium, the virial theorem yields the total mass 
\begin{equation}
M = {3\sigma^2 R\over G} \; ,
\end{equation}
where $G$ is the gravitational constant, $\sigma$ is the dispersion of the galaxy velocities along the 
line of sight, and $R$ is the cluster size 
\begin{equation}
 R = {\pi\over 2} {N(N-1)\over 2}\left(\sum_i \sum_{i>j} {1\over r_{ij}} \right)^{-1} \; ,
\label{eq:size}
\end{equation}
where $r_{ij}$ is the projected separation between galaxies $i$ and $j$, and 
$N$ is the number of galaxies; equation (\ref{eq:size}) assumes that all
the galaxies  have the same mass.
Clusters are not isolated systems and the correct application of the virial theorem
requires the inclusion of a surface term \citep{the86, girardi98}. However, 
correcting for this factor, and correcting also for departures
from spherical symmetry, for galaxies of different masses and for the presence of interlopers, namely galaxies which 
are not cluster members but appear projected onto the cluster field of view,
is not sufficient to fill in the discrepancy between the virial mass and
the sum of the galaxy masses. We can reach the same conclusion with a more
sophisticated dynamical analysis based on the Jeans equations \citep{biviano06}. 
Zwicky's result was the first indication
of the existence of dark matter. 

With the advent of astrophysics from space in 1960's, the {\it Uhuru} satellite revealed that clusters
are the most luminous extended X-ray sources on the sky  \citep[Fig.~\ref{fig:Coma}]{gursky72}.
The origin of the X-ray luminosity $L_X$ was very early interpreted as thermal Bremsstrahlung emission from 
a hot intracluster plasma (\citealt{felten66}; \citealt{bykov08b} - Chapter 8, and 
\citealt{kaastra08} - Chapter 9, this volume):
\begin{equation}
 L_X  = \int n_{\rm e}({\bf r}) n_{\rm ions}({\bf r}) \Lambda[T({\bf r})] {\rm d}^3{\bf r} \; ,
\end{equation}
where $n_{\rm e}$ and $n_{\rm ions}$ are the electron and ion number densities in the intracluster medium (ICM)
and $\Lambda(T)$ is a cooling function. 
For temperatures $kT \gtrsim 2 $ keV  ($k$ is
the Boltzmann constant), when the ICM is almost fully ionised, we have 
$\Lambda(T) \propto T^{1/2}$. 
The existence of a hot diffuse X-ray emitting gas
implies the presence of a deep gravitational potential well that keeps the
gas confined. This evidence made untenable the hypothesis that clusters are chance superpositions of galaxies
or expanding systems \citep{amba61}.
By assuming hydrostatic equilibrium and spherical symmetry, the cumulative mass within radius $r$ is 
\begin{equation}
M(<r) = -{kTr\over G\mu m_{\rm p}} \left({{\rm d}\,\ln \rho_{\rm gas}\over {\rm d}\,\ln r} + {{\rm d}\,\ln T \over {\rm d}\,\ln r} \right) \; ,
\end{equation}
where $\mu$ is the mean molecular weight, $m_{\rm p}$ the proton mass and $\rho_{\rm gas}$ the gas mass
density. This type of analysis
confirmed that at least $70\%$ of the cluster mass is dark \citep{cowie87}, and the rest
is almost all in the diffuse ICM component: galaxies only contribute a few percent to
the total cluster mass, as inferred by Zwicky in 1937. 

\begin{figure}    
\centering \includegraphics[width=\textwidth]{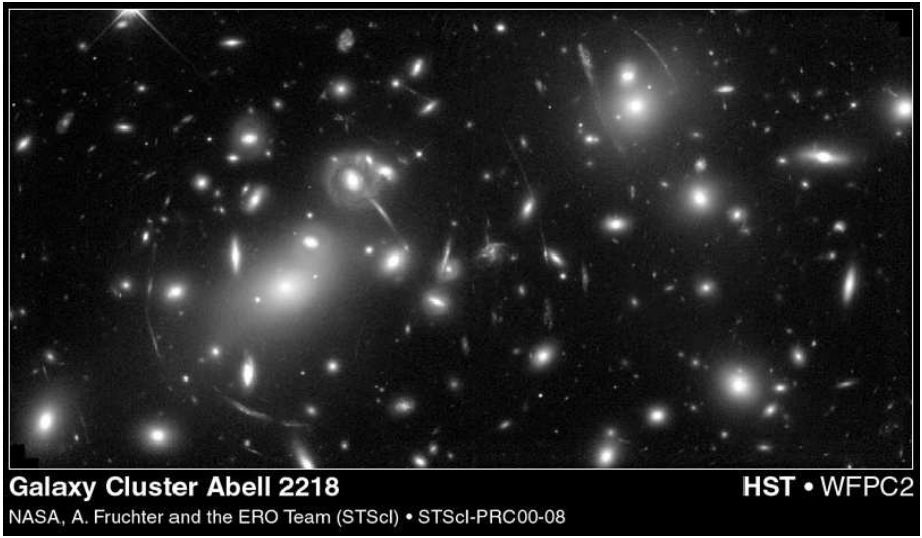}
\caption{HST optical image of Abell cluster A~2218. Many arcs and arclets are the images of background galaxies 
distorted by the gravitational field of the cluster (from http://antwrp.gsfc.nasa.gov/apod/ap011007.html).}
\label{fig:stronglens}
\end{figure}

In this same 1937 paper, Zwicky also realized that one way to avoid the assumption of dynamical
equilibrium when estimating the cluster mass is to use the deflection of light of
background sources determined by 
the cluster gravitational potential well (Fig.~\ref{fig:stronglens}); for an axially symmetric cluster, 
the total mass projected along the line of sight within the distance $r$ of closest approach of 
the light rays to the cluster centre is
\begin{equation}
 M(<r) = {rc^2\over 4G} \alpha\; ,
\end{equation}
where $c$ is the speed of light, and $\alpha$ is the deflection angle \citep[e.g.][]{schneider06}. 
The first indication of a gravitational lensing effect in 
a galaxy cluster appeared fifty years later \citep{lynds86}.
To date, both strong gravitational lensing \citep{hennawi06}, which creates
multiple images of background sources, and
weak gravitational lensing \citep{dahle07}, which produces small 
induced ellipticity in the shape of background galaxies,
indicate that,
if General Relativity is correct on the scales of clusters, $\sim 80-90$\% of the
cluster mass is dark. 

An alternative method to estimate the cluster
mass, both in their virial region and their outskirts, which does not
rely on the virial equilibrium assumption, is the caustic technique
introduced by \citet{diaferio97} and \citet{diaferio99}. 
This technique exploits the distribution of galaxies in redshift space
to infer the galaxy escape velocity from the gravitational potential 
well of the cluster, and hence its mass. In fact, in the redshift diagram, namely
line-of-sight velocity vs. projected distance $r$ from the cluster centre,
cluster galaxies populate a trumpet-shaped region limited by two curves, named 
caustics, whose separation ${\cal A}(r)$ at each radius $r$ is proportional 
to the escape velocity.
For a spherically symmetric cluster, the mass within radius $r$ is
\begin{equation}
M(<r) = {1\over 2G} \int_0^r {\cal A}^2(x) {\rm d}x \; . 
\end{equation}

All the mass estimation methods used to date, and described here, indicate that the dark 
matter contributes $\sim 80-90$\% of the total cluster mass, the ICM
contributes $\sim 10-20$\%, and the galaxies contribute less than a few percent.
With the advance of X-ray spectroscopy, it has become clear that the ICM is chemically
enriched to $\sim 0.5$ Solar abundances (\citealt{werner08} - Chapter 16, this volume): given the
relative mass contribution of the cluster components, it follows that the total mass in metals in the ICM
is larger than the sum of the mass in metals of the individual galaxies.

\subsubsection{Clusters as probes for cosmology}

\begin{figure}    
\centering \includegraphics[width=\textwidth]{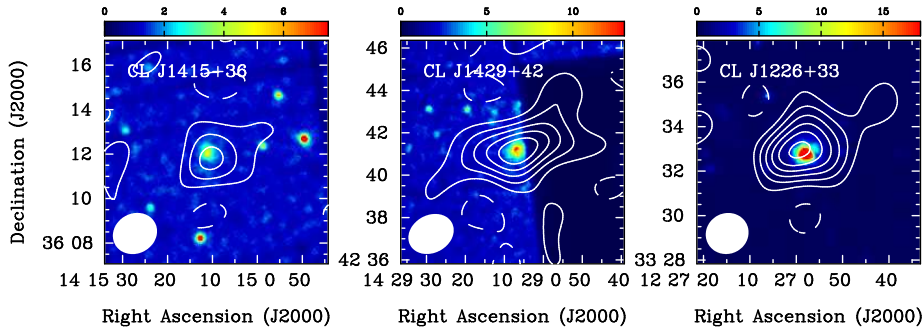}
\caption{Contours of the SZ effect measurements with SZA overlaid on the X-ray images (XMM EPIC/MOS)
of the clusters ClJ1415.1+3612 ($z=1.03$), ClJ1429.0+4241 ($z=0.92$) and ClJ1226.9+3332 ($z=0.89$). The filled
ellipses show the FWHM of the SZ observations. See \citet{muchovej07} for details.}
\label{fig:SZcluster}
\end{figure}

Right after the detection of the X-ray emission from clusters, \citet{sunyaev72}
realised that inverse Compton scattering of the Cosmic Microwave Background
(CMB) photons by the free electrons of the ICM can produce secondary
anisotropies in the CMB maps: ${\Delta I_{\rm CMB}/ I_{\rm CMB}}=f(\nu) y$,
where $f(\nu)$ is a spectral function and 
\begin{equation}
y =  \int {kT\over m_{\rm e} c^2} \sigma_{\rm T} n_{\rm e} {\rm d}l 
\label{eq:Compton-y}
\end{equation}
is the Comptonization parameter; here, $m_{\rm e}$ is the electron mass, $\sigma_{\rm T}$ the 
Thomson cross-section, and the integral is along the line of sight $l$. 
Typical ICM temperatures $kT$ and electron number densities $n_{\rm e}$ in clusters yield
$\Delta I_{\rm CMB}/I_{\rm CMB}\approx 10^{-4}$ (Fig.~\ref{fig:SZcluster}). 
Only over the last few years the sensitivity of radio telescopes
and the control of systematics have been sufficient to measure reliably this Sunyaev-Zeldovich (SZ) effect 
and its spectral signature in massive clusters \citep{bonamente06}. 
The major advantage of the SZ effect is that
the amplitude of the CMB fluctuation is independent of the cluster
redshift $z$, because the CMB intensity increases as $(1+z)^4$ and compensates
the $(1+z)^{-4}$ decrease of the SZ ``surface brightness''. 
This property has recently boosted an entire research field of cluster
surveys with the SZ effect \citep{carlstrom02}. In fact, "flux-limited" SZ cluster surveys are almost 
mass-limited cluster surveys \citep{bartlett06}. This property is
extremely relevant, because the mass of clusters is a sensible
cosmological probe, as we explain below.

If clusters are in virial equilibrium, we can derive simple relations between
their global properties, namely mass, galaxy velocity dispersion, number of
galaxies (richness), X-ray luminosity, ICM temperature, and so on. 
By considering for example the virial relation $3 kT/(2 \mu m_{\rm p})= GM/R$, 
the scaling relation between the total mass $M$ and the gas temperature $kT$ reads 
\begin{equation}
kT = 3.229 \left(\mu\over 0.6\right) 
\left(\delta\over 500\right)^{1/3} \left(M\over 10^{14} h^{-1} {\rm M}_\odot\right)^{2/3} {\rm keV}
\label{eq:M-T}
\end{equation}
where $\delta$ is the average cluster overdensity
w.r.t. the critical density $3H_0^2/(8\pi G)$ of the universe, with $H_0=100 h$ km s$^{-1}$ Mpc$^{-1}$ 
being the Hubble constant at the present time. Quantities in equation (\ref{eq:M-T}) are normalized
to typical cluster values. Analogously, we can write the total 
X-ray luminosity 
\begin{eqnarray}
L_X & = & 1.327 \times 10^{43} \left(f_{\rm gas}\over 0.1 h^{-3/2}\right)^2 \left(0.6\over \mu\right) \left(n \over 10^{-3} h^2\, {\rm cm}^{-3}\right) \left( T\over {\rm keV}\right)^{0.4}\times \cr 
&\times & \left(M\over 10^{14} h^{-1} {\rm M}_\odot\right) h^{-2}\, {\rm erg\, s}^{-1} \; ,
\end{eqnarray}
where we approximated the cooling function $\Lambda(T) = 0.843\times 10^{-23}(kT/{\rm keV})^{0.4}$
erg cm$^3$ s$^{-1}$ at $kT\gtrsim 1$ keV, which holds for gas with poor
metallicity, and assumed $n_{\rm e}=n_{\rm ions}\equiv n=f_{\rm gas}\rho/(\mu m_{\rm p})$, where
$f_{\rm gas}$ is the fraction of the cluster total mass in the ICM and $\rho$ is the 
cluster total mass density.

\begin{figure}
\centering 
\includegraphics[width=\textwidth]{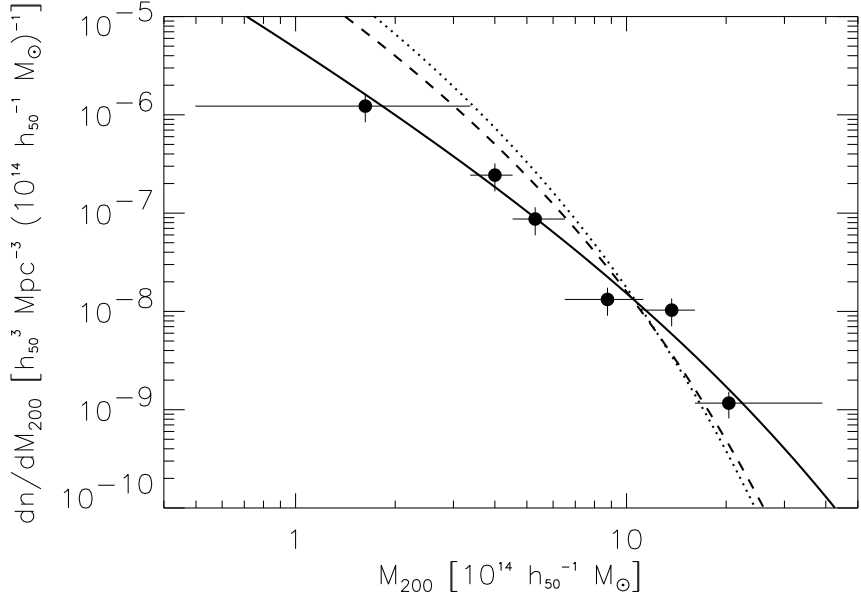}
\caption{Mass function of the HIFLUGCS X-ray clusters (dots with error bars). The solid line
is the best fit with $\Omega_{\rm m}=0.12$ and $\sigma_8=0.98$. 
The dashed and dotted lines are the best fits with $\Omega_{\rm m}=0.5$ or $\Omega_{\rm m}=1.0$ 
held fixed, which yield $\sigma_8=0.60$ and $\sigma_8=0.46$, respectively. From \citet{reiprich02}. }
\label{fig:cluster-mass-function}
\end{figure}

These scaling relations provide
a straightforward method to estimate the cluster mass from a directly observable quantity, 
like the X-ray temperature.
Quantities related to X-ray observations have received particular attention 
because of their robustness and the relatively simple physics involved 
in their correlations \citep{rosati02}.

Cluster masses are extremely relevant to constrain the cosmological model, because
clusters populate the exponential tail of the mass function of galaxy systems.
If the power spectrum of the primordial perturbations of the mass
density field in the early universe is a power law with index $n$, the
number per unit volume $[{\rm d}n(M)/{\rm d}M]dM$ of galaxy systems with total mass in the range ($M$, $M+dM$) is 
\citep{press74}
\begin{equation}
{{\rm d}n(M)\over {\rm d}M} {\rm d}M = {1\over \sqrt{\pi}} {\bar\rho\over M^2} \left(1+{n\over 3}\right)\left(M\over M_*\right)^{(n+3)/6}
\exp\left[-\left(M\over M_*\right)^{(n+3)/3}\right]{\rm d}M \; ,
\label{eq:PressSchechter}
\end{equation}
where $\bar\rho$ is the (constant) comoving mean mass density of the universe and
$M_*$ is a parameter depending on the normalisation
of the power spectrum $\sigma_8$ and on the structure growth factor, which, in turn, depends on time, the 
cosmological density parameter $\Omega_m$ and the cosmological constant $\Omega_\Lambda$. 
$M_*$ increases with time when $n>-3$ and $M_*\sim  10^{14} h^{-1}$ M$_\odot$
at the present epoch. Clearly, since the exponential cut off dominates the mass function at $M\gtrsim M_*$, the
evolution of the cluster number density is a very sensitive indicator of the power spectrum normalisation
and of the cosmological parameters (Fig.~\ref{fig:cluster-mass-function}). 
In practice, the application of this idea requires 
modern versions of the Press-Schechter mass function which 
are more sophisticated than eq. (\ref{eq:PressSchechter}) \citep{sheth99}; moreover, 
care must be taken to deal with the degeneracy among the cosmological parameters \citep[e.g.][]{voit05},
especially between $\Omega_{\rm m}$ and $\sigma_8$, which follow a relation of the form $\sigma_8=A\Omega_{\rm m}^{-\alpha}$,
with $A\approx 0.5$ and $\alpha\approx 0.5$ \citep[e.g.][]{reiprich02}. 
Nevertheless, by simply assuming that the mean mass density of the universe must
be smaller than the mean density within clusters and larger than the mean density
computed including only the richest clusters, Abell already in 1965 estimated $\Omega_{\rm m}\approx 0.2$,
a value remarkably close to the currently accepted estimate $\Omega_{\rm m}=0.26$ \citep{spergel07}.   

In most mass estimation techniques, the assumption of dynamical equilibrium is fundamental
for obtaining accurate estimates. There is however overwhelming evidence that many clusters
are out of equilibrium, as described in the following sections.

\subsection{Evidence of non-equilibrium and formation processes}
\label{subsec:noneq}

\subsubsection{Cluster galaxies}

\begin{figure}
\centering \includegraphics[width=0.7\textwidth,angle=90]{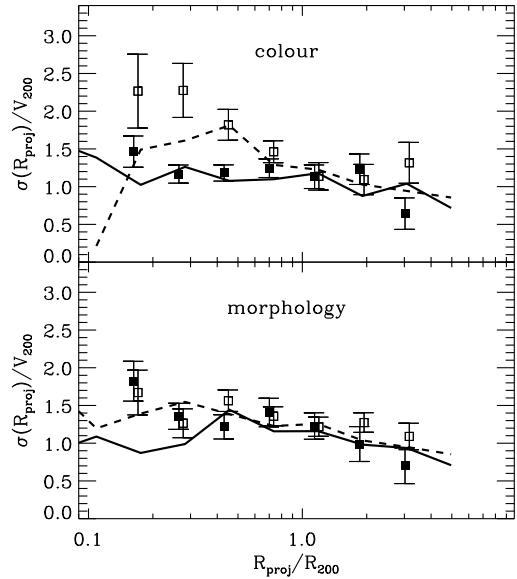}
\caption{Differential velocity dispersion profiles of galaxies
brighter than $M_R=-20.5$  in the CNOC1 clusters ($0.18<z<0.55$).
{\it Top panel}: Profiles of blue (open squares) and red (filled squares) galaxies;
the rest-frame colour threshold is $B-V=0.85$. The dashed
and the solid lines show the corresponding
profiles of the  blue and red galaxies in mock clusters 
simulated with the N-body+semi-analytic technique (Diaferio et al. 2001; see Sect.~\ref{sec:mods}).
{\it Bottom panel}: Velocity dispersion profiles of disk-dominated galaxies (open squares) and 
bulge-dominated galaxies (filled squares). 
See Diaferio et al. (2001) for details.}
\label{fig:blue-red-sigma}
\end{figure}

Elliptical and lenticulars (early-type) galaxies tend to reside in high-density regions, whereas
spiral (late-type) galaxies are more common in low-density regions. This morphology-den\-sity
relation has been known since a publication of Shapley in 1926, and possibly earlier. 
The first statistical analysis based on a large sample of 55 clusters 
showed that, in the local universe, the fraction of spirals is 80\%, 60\% and 0\%
in the field, outskirts and cores of clusters, respectively \citep{dressler80}. Subsequent work
confirmed the connection between galaxy properties and local density:
spirals in clusters are \ion{H}{i}- and dust-deficient, have larger metallicity and 
show an excess of radio continuum emission \citep[see][for a review]{boselli06}. 

At higher redshifts ($z\sim 1$), the fraction of spirals and irregulars 
in high-density regions increases at the expense of the fraction of lenticulars
\citep{postman05, smith05}. Larger gas reservoirs and galaxy close encounters and collisions, which
are more frequent in dense environments and favour the transformation of galaxy morphology, increase
the star formation activity, as it occurs in starburst galaxies;
 indeed mid-infrared observations of luminous infrared galaxies
show a more intense star formation activity in cluster environment at increasing redshift 
\citep{metcalfe05, bai07}.
Although, on average, the properties of galaxies
vary smoothly from the cluster centre to the outskirts, before reaching
the average properties of galaxies in the field \citep[e.g.][]{rines05}, 
the galaxy properties within individual clusters can show significant spatial variations, like
the different slopes of the galaxy luminosity function 
in different regions of Coma \citep{adami07}. 

In clusters with irregular morphology, the major axes 
of the bright galaxies tend to be aligned 
both with each other and with the major axis of the parent cluster \citep{plionis03}, 
as a century ago \citet{wolf02} already pointed out for the galaxies in Coma. 
An alignment between the major axis of the bright central galaxy (BCG) or the
cD galaxy and the major axis of the parent cluster is also present, as first noticed by
\citet{carter80}. Any alignment present in the initial conditions is expected
to be erased by multiple galaxy encounters occurring in the 
cluster environment. Therefore, the presence of alignment should indicate
a young dynamical state of the cluster.  

Blue cluster galaxies show larger velocity dispersions than 
red galaxies \citep[Fig.~\ref{fig:blue-red-sigma};][]{biviano97}. This difference is
attributed to more elongated orbits \citep{biviano04}, indicating that,
unlike the red galaxies which are likely to be in
dynamical equilibrium within the cluster gravitational potential well, blue galaxies are,
on average, possibly falling onto the cluster for the first time.

Intracluster stars detected with planetary nebulae or as diffuse light 
in clusters at low \citep{gerhard05} and intermediate redshift \citep{zibetti05} can be the relics of tidal
interactions suffered by falling galaxies \citep{covone06} and/or the
relics of merging of galaxies already in place in the forming gravitational
potential well of the cluster \citep{murante07}, as observed in real
clusters \citep{rines07}; intracluster stars can contain from 10 to 50\%
of the mass in stars in the core of clusters and can show distinct kinematic substructure
\citep{arnaboldi06}.

\subsubsection{Substructure, mass accretion and non-thermal phenomena}

\begin{figure}
\begin{center}
\hbox{
\includegraphics[width=5.8cm]{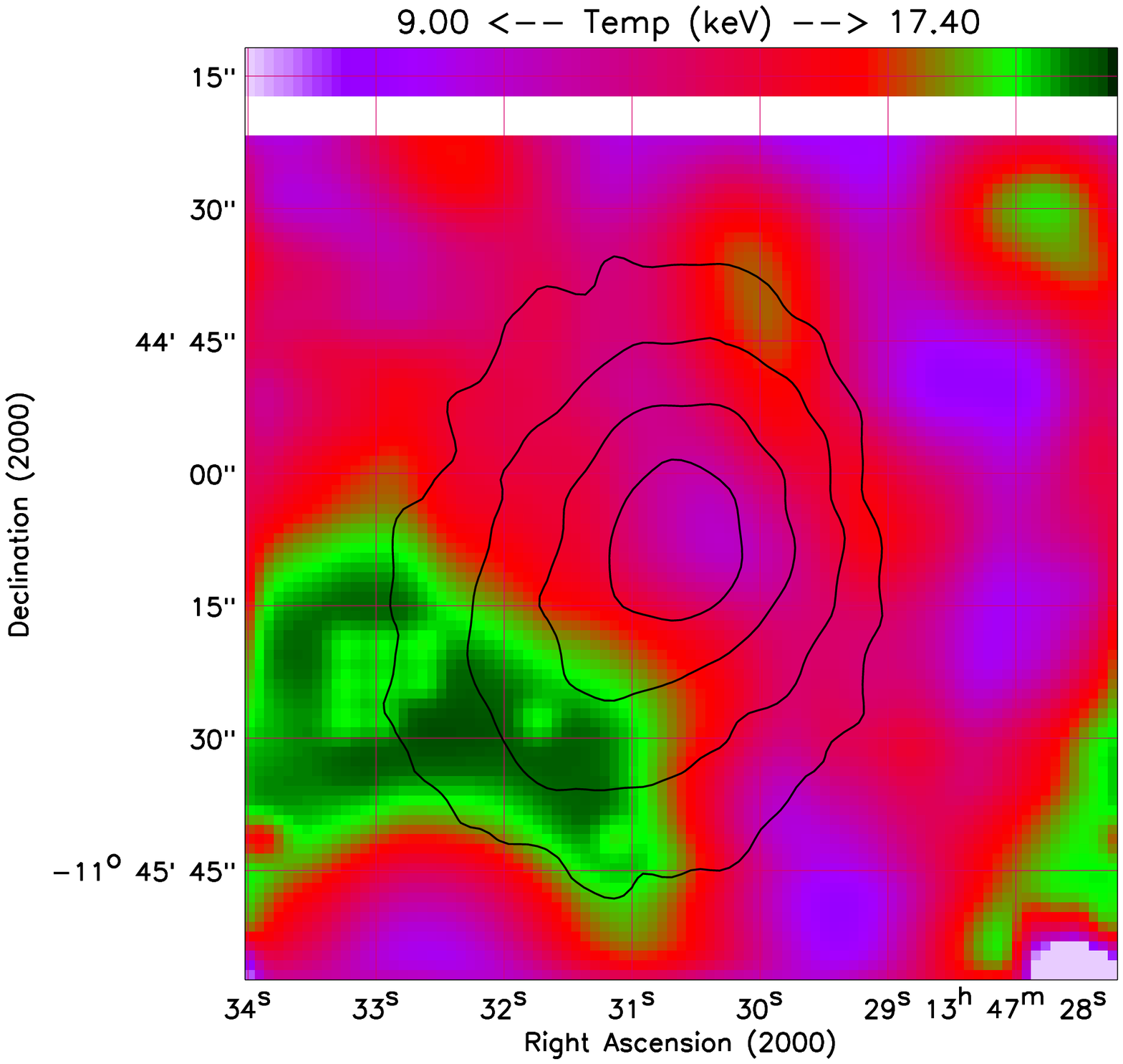}
\includegraphics[width=6.0cm]{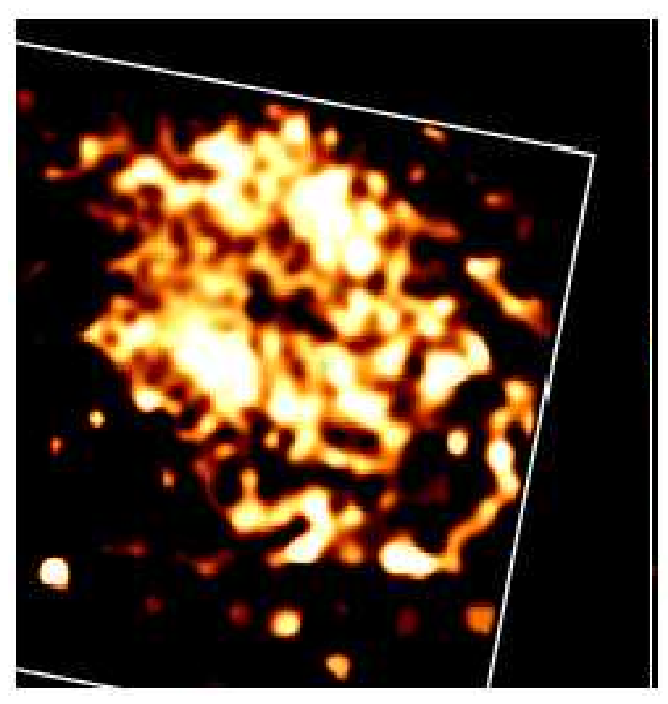}
}
\caption{{\it Left panel}: Temperature map of the cluster RX J1347.5-1145. The superposed contours
show the X-ray surface brightness. See \citet{gitti04} for details.
{\it Right panel}: Map of the metallicity abundance of the central region of A426, 
the Perseus cluster; brighter regions
are more metal rich. See \citet{sanders05} for details.}
\label{fig:substructure}
\end{center}
\end{figure}

A large fraction of clusters shows the presence of substructure
both in their galaxy distribution and in their X-ray emission morphology. The fraction 
of clusters with substructure depends
on the cluster sample and on the substructure identification technique,
but it is substantial, lying in the range 30-80\% \citep[see][and references therein]{ramella07}.
The X-ray band also shows patchy temperature \citep{belsole05} and metallicity
maps \citep[e.g.][Fig.~\ref{fig:substructure}; 
see \citealt{werner08} - Chapter 16, this volume]{hayakawa06, finoguenov06}. 
Moreover, the X-ray morphology is increasingly irregular 
with increasing redshift \citep{jeltema05}.
Where the angular resolution is low and detailed maps of the X-ray surface brightness
cannot be obtained, a clear indication that, in some clusters, the ICM might be out of equilibrium
comes from the hard, possibly thermal, excess (\citealt{rephaeli08} - Chapter 5, this volume),  
which appears when the X-ray spectrum is fitted assuming a plasma with single
temperature and metallicity. In these cases, 
two or more temperatures are required to yield a reasonable fit.

All these pieces of evidence suggest that clusters accrete matter
from the surrounding regions. The cluster 1ES~0657-55.8, dubbed ``the bullet cluster'' \citep{markevitch02, barrena02}, is a spectacular example
of a merging cluster at intermediate redshift ($z=0.296$). At low redshifts, there are many other examples
of merging clusters or galaxy groups falling
onto clusters along filamentary structure. To mention a few: 
Coma \citep{colless96, adami05}, A~521 \citep{ferrari03}, A~754 \citep{henry04}, A~2199 \citep{rines01}, 
A~2219 \citep{boschin04}, A~3560 in the Shapley supercluster \citep{bardelli02}, A~3921 \citep{ferrari05}. 
At high redshift the number of irregular and merging clusters increases substantially \citep{rosati04, kodama06}. 

Accretion of matter is a natural explanation for the
patchy morphology of the X-ray emission and metallicity maps; it can also cause most 
of the turbulence in the ICM \citep{schuecker04} and generate
shocks (\citealt{bykov08a} - Chapter 7, this volume), 
which can be efficient generators of high-energy 
cosmic rays. 

Non-thermal processes in the 
ICM appear in the X-ray and radio bands (\citealt{rephaeli08} - Chapter 5, this volume). 
Non-thermal hard X-ray emission
is indeed observed in a few systems \citep[e.g.][and references therein]{petrosian06, 
fusco-femiano07}.  However, the existence of both the hard X-ray 
emission excess \citep{rossetti04} and the soft X-ray and UV emission
excesses, regardless of their thermal or non-thermal origin, are still debated 
\citep[e.g.][- Chapter 4, this volume]{lieu05, nevalainen07, werner07,
durret08}.

At radio frequencies, the cluster emission can be both extended and
associated with galaxies. One can infer the direction of the bulk motion 
of gas clouds within the ICM from the deviations that the ICM kinetic pressure
exerts on the radio jets emerging from active galactic nuclei (AGN) \citep{burns02}.

The extended radio emission has a steep spectrum and can appear with two different morphologies:
either the emission comes from the cluster centre (halo), is 
extended and regular, or the emission comes from the cluster outskirts and is elongated (relic)
(\citealt{feretti07}; \citealt{ferrari08} - Chapter 6, this volume). 
By combined radio and hard X-ray observations or by Faraday rotation measurements
one can infer the presence of magnetic fields in the ICM with intensity ranging from a few tenths to several $\mu$Gauss 
\citep[e.g.][- Chapter 6, this volume]{blasi07, ferrari08}. 

Both the radio and the hard X-ray emissions originate from relativistic electrons:
the former is synchrotron radiation, the latter is emission due to inverse Compton scattering with the CMB
photons [\citealt{petrosian08a} - Chapter 10, this volume]. 
Among the many acceleration mechanisms proposed 
to produce the relativistic electrons [\citealt{petrosian08b} - Chapter 11, this volume], 
shocks due to the accretion of matter from the surrounding regions seem to be
one of the most efficient.

\subsection{Connection with the large-scale structure}
\label{subsec:LSSconn}

\subsubsection{A morphology-density relation for clusters}

It has been known for a long time that clusters are not randomly distributed in the universe but form large
concentrations named superclusters \citep{shapley33}.
The connection between clusters and the large-scale structure
has become evident with extended galaxy redshift surveys, like the CfA \citep{delapparent86},
the SDSS \citep{york00}, and the 2dF \citep{colless01} redshift surveys.

\begin{figure}
\centering \includegraphics[width=7.cm]{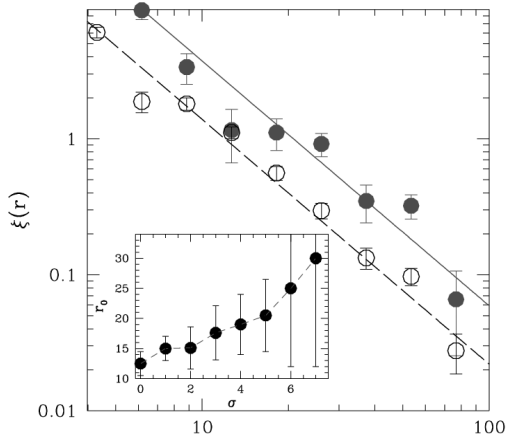}
\caption{Two-point correlation function of the 903 APM cluster sample (open circles) and of a subsample
of clusters with significant substructure (filled circles). Insert: Correlation length $r_0$ vs. the
significance level $\sigma$ of substructure. See \citet{plionis02} for details. 
}
\label{fig:csi-subs}
\end{figure}

To quantify the distribution of objects in space, the simplest
quantity to compute is the two-point correlation function $\xi(r)$, which yields the
probability of finding two objects at separation $r$ in excess to
a Poisson distribution. Clusters correlate according to the
correlation function $\xi=(r/r_0)^{\gamma}$, with $\gamma \approx -1.8$ \citep{bahcall83}. The 
correlation length $r_0$ is proportional to the cluster richness:
richer clusters are rarer and their mean intercluster separation $d$ larger.
One finds $r_0 \propto d^{0.5}$, and $r_0\approx 20-25 h^{-1}$ Mpc for rich
clusters. However, this $r_0-d$ relation cannot be fully responsible for the
large range, from $\sim 10$ to $\sim 30 h^{-1}$ Mpc, covered by the correlation length $r_0$ 
when one considers clusters with increasing substructure \citep[Fig.~\ref{fig:csi-subs};][]{plionis02}.
Moreover, cluster shapes are generally elongated, 
as first quantified by \citet{sastry68}, and close clusters
have their major axes aligned \citep{binggeli82}. In the APM cluster sample, containing
903 objects, this alignment
is stronger when the fraction of substructure is larger \citep{plionis02}.

This connection between the properties of clusters and their large-scale environment
is also apparent in the X-ray band: the X-ray luminosity of clusters \citep{bohringer04}
and the irregularity of the X-ray surface brightness maps \citep{schuecker01} are larger in environments with a higher cluster number density.

These features appear in a scenario \citep{west94} where clusters form by episodic mass accretion along preferential directions
\citep{colberg99}, in agreement with the currently accepted model of the formation
of cosmic structure (see Sect.~\ref{sec:mods}).
Future estimates of cluster peculiar velocities based on the kinetic SZ effect,
when corrected of their systematics \citep{diaferio05a}, 
will further test whether the correlation between 
the evolution of the peculiar
velocity and the local density predicted by this model \citep{diaferio00, sheth01} is correct.

\subsubsection{The surrounding WHIM}

The optical depth of the intergalactic medium at $z\gtrsim 2$ derived from the Lyman-$\alpha$ forest
of QSO spectra \citep{rauch97}, 
the abundances of light elements combined with
the predictions of the standard Big Bang nucleosynthesis model and the power spectrum 
of the CMB \citep{spergel07} are all consistent with a
density of baryonic matter $\Omega_{\rm b}h^2 \simeq 0.022$. At low redshift, baryonic matter
within galaxies and within the diffuse hot medium of galaxy clusters and groups
only accounts for $\sim 15$\% of this $\Omega_{\rm b} $ \citep{fukugita98}.

The hierarchical clustering model solves this discrepancy by predicting that
$50\%$ of the baryons permeating the matter distribution is in a
warm phase with $kT$ in the range $0.01-1$ keV, overdensity in the range $0.1-10^4$ 
and median $\sim 10-20$, and median metallicity $\sim 0.2 $ Solar, 
with a large spread, for oxygen \citep[][- Chapter 14, this volume]{cen99, cen06, schaye08}. 
The WHIM overdensity increases with increasing dark matter
overdensity. Therefore, the outskirts of clusters and their environments are expected 
to be sources of a thermal soft X-ray emission. However, both the
expected low surface brightness of the WHIM and the intergalactic and Galactic
absorption make this observation very challenging.
Nevertheless, there are claims of detection in many clusters,
including Coma \citep{takei07}.
In fact, rather than detecting the thermal X-ray emission (\citealt{durret08} - Chapter 4, this
volume) directly, the most promising
technique is searching for thermal UV and
X-ray absorption lines in the spectra of bright background quasars or blazars (\citealt{richter08} - Chapter 3, 
this volume).
Even with this strategy, current instruments (\citealt{paerels08} - Chapter 19, this volume) 
must be used
at their sensitivity limits 
and the most plausible detections \citep[e.g.][]{nicastro05} 
are indeed still debated \citep{kaastra06}. 
Firm detections are however  of great relevance, 
because the existence of the WHIM and the estimate of its properties 
would be an extraordinary confirmation of the
current model of structure formation.

\section{The modeling framework}
\label{sec:mods}

\subsection{Players: gravity and beyond}
\label{subsec:players}

Modeling the formation and evolution of galaxy clusters requires the set-up of 
a complete cosmological context. According to the current paradigm 
of structure formation \citep[as reviewed, e.g., in][]{springel06}, 
inflation amplifies quantum fluctuations, which are present
in the dark matter density field of the early universe, to cosmic scales; 
at the time of matter and radiation
density equality, these fluctuations start growing by gravitational instability.
After radiation and ordinary baryonic matter decouple, 
radiation pressure stops supporting the baryonic density perturbations
against self-gravity and against the pull of the dark matter
gravitational potential wells that have already formed: the dark matter 
inhomogeneities thus accelerate the collapse of ordinary matter.

\begin{figure}
\centering \includegraphics[width=7.cm]{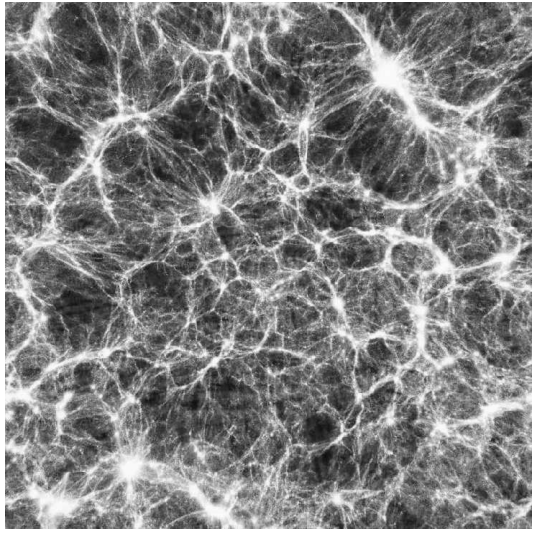}
\caption{Map density at $z=0$ of the gas in a slice of the N-body/hydrodynamics simulation of Borgani et al. (2004).
The volume shown is $192 h^{-1}$ Mpc on a side and $12 h^{-1}$ Mpc thick. The
high-density regions are white, the low-density regions dark. See \citet{borgani04} for details.}
\label{fig:nbodysim}
\end{figure}

The pre-inflationary Gaussian quantum fluctuations generate a scale-free power spectrum of 
the primordial density perturbations. At later times, the dark matter particles, which have 
sufficient momenta to stream out of the denser
regions, set in a characteristic scale in the power spectrum, which is 
proportional to the typical velocity of the dark matter particles 
themselves; below this scale the density perturbations
are damped. The existence of quasars and 
galaxies at high redshift \citep{fan04}, the power spectrum of the CMB 
\citep{smoot92,debernardis00,hanany00}, and 
the power spectrum of the large-scale distribution of galaxies 
\citep{tegmark04} all agree with the hypothesis that the dark matter is ``cold'', namely
that the dark matter particles become non-relativistic at very
early times and the damping scale is $\ll 1h^{-1}$ Mpc. In these Cold Dark Matter (CDM) models, larger
systems form by the aggregation of smaller systems and galaxy clusters
are the nodes of the network of sheets and filaments which constitutes the matter density field 
(Fig.~\ref{fig:nbodysim}). 
Clusters can form at reasonably recent times, with the actual formation time depending on
the cosmological parameters.

In CDM models, the scale-free properties of the primordial power
spectrum are preserved on scales larger than the scales of galaxies. 
This feature implies that the dark matter halos of galaxies, groups and
clusters evolve self-similarly. If gravitation is the only force playing, 
the physics of the ICM is ruled by adiabatic compression and shock heating, 
and we can derive simple relations between cluster properties 
at different redshifts \citep{kaiser86}. The cluster virial mass scales as
$M_{\rm vir} \propto \rho_{\rm c}(z) \Delta_{\rm c}(z) R^3 $, where $R$ is the cluster
size in physical units, $\rho_{\rm c}(z)=3H^2(z)/(8\pi G)$ is the critical
density of the universe, and $\Delta_{\rm c}(z)$ is the cluster density in units of $\rho_{\rm c}(z)$;
\footnote{A well-known approximation is 
\begin{equation}
\Delta_{\rm c}(z)=18\pi^2 + \left\{ \begin{array}{ll}
 60 w - 32 w^2, &\Omega_{\rm m}\le 1,\quad \Omega_\Lambda= 0\\
82 w - 39 w^2, &\Omega_{\rm m}+\Omega_\Lambda=1
 \end{array}\right.
\label{eq:Dcfit}
\end{equation}
where $w=\Omega_{\rm m}(z)-1$ \citep{bryan98}.} 
$\rho_{\rm c}(z)$ scales with redshift $z$ as $\rho_{\rm c}(z)\propto E^2(z)
= \Omega_{\rm m}(1+z)^3 + (1-\Omega_{\rm m}-\Omega_\Lambda)(1+z)^2+ \Omega_\Lambda$. 
The cluster size thus scales as $R\propto M_{\rm vir}^{1/3} \Delta_{\rm c}^{-1/3}(z) E^{-2/3}(z)$,
and the temperature as $T\propto M_{\rm vir}/R \propto M_{\rm vir}^{2/3} \Delta_{\rm c}^{1/3}(z) E^{2/3}(z)$; thus,
for example, one finds for the X-ray luminosity due to Bremsstrahlung emission:
\begin{equation}
L_X\propto \int n_{\rm e} n_{\rm ions} \Lambda(T) {\rm d}V \propto M_{\rm vir}^2 R^{-3} T^{1/2} \propto M_{\rm vir}^{4/3}
 E^{7/3}(z)\Delta_{\rm c}^{7/6}(z)
\end{equation}
or for the Comptonization parameter
\begin{equation}
y\propto \int n_{\rm e} T {\rm d}l \propto M_{\rm vir} R^{-2} T \propto M_{\rm vir} E^2(z)\Delta_{\rm c}(z) \; .
\end{equation}

These scaling relations only partially agree with
observations; in fact, (1) the relative contribution of 
non-gra\-vi\-ta\-ti\-onal sources (AGN, galactic winds, supernovae, 
turbulence) to the ICM energy budget depends on the mass of the cluster and affects the slope of the scaling 
relations; and (2) the merging history of the cluster introduces an intrinsic scatter
in the relations \citep{cavaliere98}. 
Moreover, clusters at high redshift have scaling relations steeper
than the self-similar prediction \citep{ettori04}.

The deviations from the self-similar scaling relations are clearly related to non-gra\-vi\-ta\-ti\-onal
processes. The interplay between gravitational and non-gravitational processes is highly
non-linear and treating it appropriately requires sophisticated modeling.
Over the last thirty years two main approaches have been followed: numerical
simulations (\citealt{dolag08a} - Chapter 12, this volume) and an analytic (or semi-analytic) approach. Over the last ten years, 
the two approaches have been combined. 

\subsection{Making clusters: recipes, ingredients and tools}
\label{subsec:recipes}

For the last thirty years, numerical simulations have
represented the main tool of theoretical astrophysicists 
to investigate the non-linear evolution of cosmic structure and compare
the theoretical results with observations.
The first computational device for simulating galaxy interactions used an ingenious
system of light-bulbs \citep{holmberg41}. Numerical simulations of 
astrophysical systems have now reached tremendous sophistication (\citealt{dolag08a} - Chapter 12, this volume)
when compared to the first numerical integration of the equations of motion
of a few stars in a star cluster \citep{hoerner60}.
Current numerical models reproduce the gross features of the large-scale
structure, and the distribution of galaxies and diffuse baryons (\citealt{borgani08a} - Chapter 13; 
\citealt{schaye08} - Chapter 14, this volume), 
including the statistical properties of the Ly$\alpha$ forest \citep{springel06}. 

Until the 1980's, numerical simulations relied on $N$-body techniques only
to follow the evolution of the dissipationless 
component of cosmic structure. Particles in the simulations
were first identified with galaxies and later on with dark
matter particle tracers. \citet{peebles70} and
\citet{white76} were the first to simulate the formation
of a galaxy cluster with vacuum boundary conditions. 
To obtain a realistic distribution of matter, however, it is 
necessary to simulate a cosmological volume, 
with periodic boundary conditions, which is large enough to be representative of
the universe. The first simulation of this kind,
with a cold dark matter power spectrum of the initial density perturbations,
demonstrated that dark matter must
be constituted of cold collisionless particles, rather than massive neutrinos \citep{davis85}.
Recent dark matter only simulations yield clusters whose number density
evolution \citep{evrard02}, virial relations \citep{evrard07}, and connection
between their shape and the large-scale structure
\citep{altay06, lee07} resemble the observations and can be
used as probes of the cosmological model.

In the late 1980s, smoothed particle hydrodynamics (SPH) \citep{evrard88,hernquist89} 
and mesh techniques \citep{cen90} were
introduced to follow the evolution of the dissipative component. 
The presence of dissipative gas substantially increases  
the number of physical processes to be taken into account and therefore increases the 
complexity of the simulations. In fact, {\it a priori}, it is not guaranteed that the different
techniques yield similar results and some code comparisons have
been performed over the years to check the consistency of the
various integration schemes \citep{kang94, frenk99, oshea05}. 

The simplest approximation is to consider
the gas non-radiative and perform adiabatic simulations \citep{evrard90}. This
model is able to reproduce the general trend of the self-similar scaling laws,
but disagrees with observations in detail, yielding, in general, shallower X-ray luminosity-temperature
and mass-temperature relations \citep{navarro95b}. 
To reproduce the observations more closely, and explain the deviations from the self-similar model,
radiative cooling \citep{thomas92,katz93} and 
non-gra\-vi\-ta\-ti\-onal heating processes \citep{navarro95b} must be included.
More realistic models, attempting to reproduce
the thermal properties of the ICM and the stellar content of clusters,
have to include explicitly all those intimately coupled processes which rule 
galaxy formation and the thermodynamics of the ICM: star formation, 
energy feedback from supernovae explosion \citep{valda02} and galactic winds 
(\citealt{metzler94}; \citealt{dolag08a} - Chapter 12; \citealt{borgani08a} - 
Chapter 13, this volume), besides 
metal production and chemical enrichment (\citealt{valda03}; \citealt{schindler08} - Chapter 17; \citealt{borgani08b} - Chapter 
18, this volume), which strongly affects
the gas cooling rate; moreover, one needs to include magnetic fields and its associated
non-thermal processes (\citealt{dolag99}; \cite{dolag08b} - Chapter 15, this volume). 

The galaxy population in the simulated clusters have shown some
rough similarities with observations since the very first numerical experiments
\citep{evrard94}; many other observed properties are reproduced at different levels of agreement, 
including the soft X-ray thermal emission \citep{cheng05}.
Nevertheless, in no simulation today the thermodynamical properties of the ICM, the stellar
mass fraction and the galaxy luminosity function simultaneously match the 
observations. One of the most serious discrepancies with observations is
the overproduction of stars: the fraction of baryons in stars usually is $\sim 20\%$,
at least a factor of two larger than observed \citep{borgani04}.
A possible solution has been identified in the
inhibition of the ICM cooling by the energy injection of AGN jets, as
suggested by the X-ray cavities in correspondence
of the radio emission observed around cluster central galaxies \citep{mcnamara06,
mcnamara07};
the presence of these cavities correlates with the X-ray temperature drops in 
the centre of some clusters, early interpreted as cooling flows \citep{dunn06}; 
however, the details of the heating mechanism by AGN jets remain uncertain \citep{voit05}. 

The main difficulty when treating most of the dissipative processes resides in the
fact that they occur on scales much smaller than the spatial resolution
of the simulation and must be approximated by phenomenological 
recipes: gas cooling, star formation and stellar evolution, which occur at scales $\ll 1$ pc, have
effects on galactic and extragalactic scales $\gg 1$ Mpc; thus the simulations should model a spatial
range of at least six orders of magnitude. These large dynamic ranges can 
be covered by neither a single numerical experiment nor experiments which
rely on the resimulation  technique, where an individual cluster extracted from a
large-scale cosmological simulation is resimulated at a much higher resolution
with appropriate boundary conditions \citep{navarro95a}.

The large number of physical processes and the large computational
resources required to run an $N$-body/hydrodynamics simulation
for a realistic modeling of cluster and galaxy formation 
have suggested a complementary 
strategy to explore more efficiently the parameter space: the combination of semi-analytic models
with merger trees of dark matter halos extracted from dissipationless N-body simulations \citep{roukema97, kauffmann99}. This approach has
reproduced the evolution and present-day spatial and kinematic distribution of galaxies
in the universe and in clusters \citep{springel05}.
The major shortcoming of this strategy is that severe uncertainties in the predictions
can be introduced by the crude simplification of the physical processes
and especially of the thermodynamics of the ICM, usually assumed to 
be isothermal and in hydrostatic equilibrium within each dark matter halo. 
Some improvements can be introduced by
describing the ICM thermal properties in terms of entropy
generation and distribution \citep{mccarthy07}, but much work remains to be done.

\begin{acknowledgements}
The authors thank ISSI (Bern) for support of the team ``Non-virialized X-ray
components in clusters of galaxies''. Partial support from 
the PRIN2006 grant ``Costituenti fondamentali dell'Uni\-ver\-so'' of
the Italian Ministry of University and Scientific Research
and from the INFN grant PD51 is gratefully acknowledged. 
The authors also acknowledge financial support by the
Austrian Science Foundation (FWF) through grants P18523-N16 and
P19300-N16, by the Tiroler Wissenschaftsfonds and through the
UniInfrastrukturprogramm 2005/06 by the BMWF.
\end{acknowledgements}


\begin{thebibliography}{}

\bibitem[\protect\citeauthoryear{Abell}{1958}]{abell58}
Abell, G.O.,  \aspjs{1958}{3}{211}

\bibitem[\protect\citeauthoryear{Abell}{1965}]{abell65}
Abell, G.O., \araa{1965}{3}{1} 

\bibitem[\protect\citeauthoryear{Adami et al.}{2005}]{adami05}
Adami, C., Biviano, A., Durret, F., \& Mazure, A., \aeta{2005}{443}{17}	

\bibitem[\protect\citeauthoryear{Adami et al.}{2007}]{adami07}	
Adami, C., Durret, F., Mazure, A., et al., \aeta{2007}{462}{411}

\bibitem[\protect\citeauthoryear{Altay et al.}{2006}]{altay06}
Altay, G., Colberg, J.M., \& Croft, R.A.C., \mnras{2006}{370}{1422}	

\bibitem[\protect\citeauthoryear{Ambartsumian}{1961}]{amba61}
Ambartsumian, V.A.,  \aj{1961}{66}{536}

\bibitem[\protect\citeauthoryear{Arnaboldi et al.}{2006}]{arnaboldi06}
Arnaboldi, M., Gerhard, O., Freeman, K.C., et al., 2006, in 
 Planetary nebulae in our Galaxy and beyond, 
Proc. IAU Symp. 234, eds. 
M.J. Barlow \& R.H. M\'endez, Cambridge Univ. Press, p. 337

\bibitem[\protect\citeauthoryear{Bahcall \& Soneira}{1983}]{bahcall83}
Bahcall, N.A., \& Soneira, R.M., \aspj{1983}{270}{20}

\bibitem[\protect\citeauthoryear{Bai et al.}{2007}]{bai07}
Bai, L., Marcillac, D., Rieke, G.H., et al., \aspj{2007}{664}{181} 
 
\bibitem[\protect\citeauthoryear{Bardelli et al.}{2002}]{bardelli02}
Bardelli, S., Venturi, T., Zucca, E., et al., \aeta{2002}{396}{65}	

\bibitem[\protect\citeauthoryear{Barrena et al.}{2002}]{barrena02}
Barrena, R., Biviano, A., Ramella, M., Falco, E.E., \& Seitz, S., \aeta{2002}{386}{816}	

\bibitem[\protect\citeauthoryear{Bartlett}{2006}]{bartlett06}
Bartlett, J.G., 2006, astro-ph/0606241 

\bibitem[\protect\citeauthoryear{Belsole et al.}{2005}]{belsole05}
Belsole, E., Sauvageot, J.-L., Pratt, G.W., \& Bourdin, H., 2005, Adv. Sp. Res.,  36, 630	

\bibitem[\protect\citeauthoryear{Bertone et al.}{2008}]{schaye08}
Bertone, S., Schaye, J., \& Dolag, K., 2008, SSR, in press

\bibitem[\protect\citeauthoryear{Binggeli}{1982}]{binggeli82}
Binggeli, B., \aeta{1982}{107}{338}

\bibitem[\protect\citeauthoryear{Biviano}{1998}]{biviano98}
Biviano, A., 1998, Untangling Coma Berenices: A new vision of an old cluster, in
Proc. meeting held in Marseilles (France), June 17-20, 1997, 
ed. A. Mazure, F. Casoli, F. Durret, \& D. Gerbal, Word Scient. Pub. Co Pte Ltd, p. 1

\bibitem[\protect\citeauthoryear{Biviano}{2000}]{biviano00}
Biviano, A., 2000, From Messier to Abell: 200 years of science with galaxy clusters, in 
Constructing the Universe with clusters of galaxies, IAP 2000 meeting, 
eds. F. Durret \& D. Gerbal, astro-ph/0010409

\bibitem[\protect\citeauthoryear{Biviano}{2006}]{biviano06}
Biviano, A., 2006, in Proc. of the XLI Rencontres de Moriond, XXVI Astrophysics Moriond Meeting: From dark halos to light, eds. L. Tresse, S. Maurogordato, 
\& J. Tran Thanh Vani, Editions Frontieres

\bibitem[\protect\citeauthoryear{Biviano \& Katgert}{2004}]{biviano04}
Biviano, A., Katgert, P., \aeta{2004}{424}{779}

\bibitem[\protect\citeauthoryear{Biviano et al.}{1997}]{biviano97}
Biviano, A., Katgert, P., Mazure, A., et al., \aeta{1997}{321}{84}	

\bibitem[\protect\citeauthoryear{Blasi et al.}{2007}]{blasi07}
Blasi, P., Gabici, S., Brunetti, G., 2007, Int. J. Mod. Phys., A22, 681

\bibitem[\protect\citeauthoryear{B\"ohringer et al.}{2004}]{bohringer04}
B\"ohringer, H., 2004, The large-scale environment of groups and clusters of galaxies, 
in Outskirts of galaxy clusters: intense life in the suburbs, IAU Coll. 195, 
ed. A. Diaferio, p. 12

\bibitem[\protect\citeauthoryear{Bonamente et al.}{2006}]{bonamente06}
Bonamente, M., Joy, M.K., LaRoque, S.J., et al., \aspj{2006}{647}{25}

\bibitem[\protect\citeauthoryear{Borgani et al.}{2004}]{borgani04}
Borgani, S., Murante, G., Springel, V., et al., \mnras{2004}{348}{1078}

\bibitem[\protect\citeauthoryear{Borgani et al.}{2008a}]{borgani08a}
Borgani, S., Diaferio, A., Dolag, K., \& Schindler, S., 2008a, SSR, in press

\bibitem[\protect\citeauthoryear{Borgani et al.}{2008b}]{borgani08b}
Borgani, S., Fabjan, D., Tornatore, L., et al., 2008b, SSR, in press

\bibitem[\protect\citeauthoryear{Boschin et al.}{2004}]{boschin04}
Boschin, W., Girardi, M., Barrena, R., et al., \aeta{2004}{416}{839}

\bibitem[\protect\citeauthoryear{Boselli \& Gavazzi}{2006}]{boselli06}
Boselli, A., \& Gavazzi, G., 2006, PASP, 118, 517

\bibitem[\protect\citeauthoryear{Bryan \& Norman}{1998}]{bryan98}
Bryan, G.L., \& Norman, M.L., \aspj{1998}{495}{80}	

\bibitem[\protect\citeauthoryear{Burns et al.}{2002}]{burns02}
Burns, J.O., Loken, C., Roettiger, K., et al. 2002, New Astron. Rev.,  46, 135

\bibitem[\protect\citeauthoryear{Bykov et al.}{2008a}]{bykov08a}
Bykov, A.M., Dolag, K., \& Durret, F., 2008a, SSR, in press

\bibitem[\protect\citeauthoryear{Bykov et al.}{2008b}]{bykov08b}
Bykov, A.M., Paerels, F.B.S., \& Petrosian, V., 2008b, SSR, in press

\bibitem[\protect\citeauthoryear{Carlstrom et al.}{2002}]{carlstrom02}
Carlstrom, J.E., Holder, G.P., \& Reese, E.D., \araa{2002}{40}{643}	

\bibitem[\protect\citeauthoryear{Carter \& Metcalfe}{1980}]{carter80}
Carter, D., \& Metcalfe, N., \mnras{1980}{191}{325}

\bibitem[\protect\citeauthoryear{Cavaliere et al.}{1998}]{cavaliere98}
Cavaliere, A., Menci, N., \& Tozzi, P., \aspj{1998}{501}{493}	

\bibitem[\protect\citeauthoryear{Cen \& Ostriker}{1999}]{cen99}
Cen, R., \& Ostriker, J.P., \aspj{1999}{514}{1} 

\bibitem[\protect\citeauthoryear{Cen \& Ostriker}{2006}]{cen06}
Cen, R., \& Ostriker, J.P., \aspj{2006}{650}{560}	

\bibitem[\protect\citeauthoryear{Cen et al.}{1990}]{cen90}
Cen, R.Y., Ostriker, J.P., Jameson, A., \& Liu, F., \aspj{1990}{362}{L41}	

\bibitem[\protect\citeauthoryear{Cheng et al.}{2005}]{cheng05}
Cheng, L.-M., Borgani, S., Tozzi, P., et al., \aeta{2005}{431}{405}	

\bibitem[\protect\citeauthoryear{Colberg et al.}{1999}]{colberg99}
Colberg, J.M., White, S.D.M., Jenkins, A., \& Pearce, F.R., \mnras{1999}{308}{593}	

\bibitem[\protect\citeauthoryear{Colless et al.}{2001}]{colless01}
Colless, M., Dalton, G., Maddox, S., et al., \mnras{2001}{328}{1039}	

\bibitem[\protect\citeauthoryear{Colless \& Dunn}{1996}]{colless96}
Colless, M., \& Dunn, A.M., \aspj{1996}{458}{435}

\bibitem[\protect\citeauthoryear{Covone et al.}{2006}]{covone06}
Covone, G., Adami, C., Durret, F., et al., \aeta{2006}{460}{381}	

\bibitem[\protect\citeauthoryear{Cowie et al.}{1987}]{cowie87}
Cowie, L.L., Henriksen, M., \& Mushotzky, R., \aspj{1987}{317}{593} 

\bibitem[\protect\citeauthoryear{Dahle}{2007}]{dahle07}
Dahle, H., 2007, astro-ph/0701598 

\bibitem[\protect\citeauthoryear{Davis et al.}{1985}]{davis85}
Davis, M., Efstathiou, G., Frenk, C.S., \& White, S.D.M., \aspj{1985}{292}{371}	

\bibitem[\protect\citeauthoryear{de Bernardis et al.}{2000}]{debernardis00}
de Bernardis, P., Ade, P.A.R., Bock, J.J., et al., 2000, Nature, 404, 955

\bibitem[\protect\citeauthoryear{de Lapparent et al.}{1986}]{delapparent86}
de Lapparent, V., Geller, M.J., \& Huchra, J.P., \aspj{1986}{302}{L1}	

\bibitem[\protect\citeauthoryear{Diaferio}{1999}]{diaferio99}
Diaferio, A., \mnras{1999}{309}{610} 

\bibitem[\protect\citeauthoryear{Diaferio \& Geller}{1997}]{diaferio97}
Diaferio, A., \& Geller, M.J., \aspj{1997}{481}{633} 

\bibitem[\protect\citeauthoryear{Diaferio et al.}{2005}]{diaferio05a}
Diaferio, A., Borgani, S., Moscardini, L., et al., \mnras{2005}{356}{1477}

\bibitem[\protect\citeauthoryear{Diaferio et al.}{2001}]{diaferio01}
Diaferio, A., Kauffmann, G., Balogh, M. L., et al., \mnras{2001}{323}{999}

\bibitem[\protect\citeauthoryear{Diaferio et al.}{2000}]{diaferio00}
Diaferio, A., Sunyaev, R.A., \& Nusser, A., 2000, ApJ, 533, L71

\bibitem[\protect\citeauthoryear{Dolag et al.}{1999}]{dolag99}
Dolag, K., Bartelmann, M., \& Lesch, H., \aeta{1999}{348}{351}	

\bibitem[\protect\citeauthoryear{Dolag et al.}{2008a}]{dolag08a}
Dolag, K., Borgani, S., Schindler, S., Diaferio, A., \& Bykov, A.M., 2008a, SSR, in press

\bibitem[\protect\citeauthoryear{Dolag et al.}{2008b}]{dolag08b}
Dolag, K., Bykov, A.M., \& Diaferio, A., 2008b, SSR, in press

\bibitem[\protect\citeauthoryear{Dressler}{1980}]{dressler80}
Dressler, A., \aspj{1980}{236}{351}

\bibitem[\protect\citeauthoryear{Dunn \& Fabian}{2006}]{dunn06}
Dunn, R.J.H., \& Fabian, A.C., \mnras{2006}{373}{959}

\bibitem[\protect\citeauthoryear{Durret et al.}{2008}]{durret08}
Durret, F., Kaastra, J.S., Nevalainen, J., Ohashi, T., \& Werner, N., 2008, SSR, in press

\bibitem[\protect\citeauthoryear{Ettori et al.}{2004}]{ettori04}
Ettori, S., Tozzi, P., Borgani, S., \& Rosati, P., \aeta{2004}{417}{13}

\bibitem[\protect\citeauthoryear{Evrard}{1988}]{evrard88}
Evrard, A.E., \mnras{1988}{235}{911}	

\bibitem[\protect\citeauthoryear{Evrard}{1990}]{evrard90}
Evrard, A.E., \aspj{1990}{363}{349}	

\bibitem[\protect\citeauthoryear{Evrard et al.}{1994}]{evrard94}
Evrard, A.E., Summers, F.J., \& Davis, M., \aspj{1994}{422}{11}

\bibitem[\protect\citeauthoryear{Evrard et al.}{2002}]{evrard02}
Evrard, A.E., MacFarland, T.J., Couchman, H.M.P., et al., \aspj{2002}{573}{7}

\bibitem[\protect\citeauthoryear{Evrard et al.}{2007}]{evrard07}
Evrard, A.E., Bialek, J., Busha, M., et al., 2007, astro-ph/0702241

\bibitem[\protect\citeauthoryear{Fan et al.}{2004}]{fan04}
Fan, X., Hennawi, J.F., Richards, G.T., et al., \aj{2004}{128}{515}

\bibitem[\protect\citeauthoryear{Felten et al.}{1966}]{felten66}
Felten, J.E., Gould, R.J., Stein, W.A., \& Woolf, N.J., \aspj{1966}{146}{955}	

\bibitem[\protect\citeauthoryear{Feretti \& Giovannini}{2007}]{feretti07}
Feretti, L., Giovannini, G., 2007, in Panchromatic view of clusters of galaxies and the large-scale structure, 
Springer Lect. Notes in Phys., eds. M. Plionis, O. Lopez-Cruz, \& D. Hughes, in press (astro-ph/0703494)

\bibitem[\protect\citeauthoryear{Ferrari et al.}{2008}]{ferrari08}
Ferrari, C., Govoni, F., Schindler, S., Bykov, A., \& Rephaeli, Y., 2008, SSR, in press

\bibitem[\protect\citeauthoryear{Ferrari et al.}{2005}]{ferrari05}
Ferrari, C., Benoist, C., Maurogordato, S., Cappi, A., \& Slezak, E., \aeta{2005}{430}{19}

\bibitem[\protect\citeauthoryear{Ferrari et al.}{2003}]{ferrari03}
Ferrari, C., Maurogordato, S., Cappi, A., \& Benoist, C., \aeta{2003}{399}{813}	

\bibitem[\protect\citeauthoryear{Finoguenov et al.}{2006}]{finoguenov06}
Finoguenov, A., Henriksen, M.J., Miniati, F., Briel, U.G., \& Jones, C., \aspj{2006}{643}{790}	

\bibitem[\protect\citeauthoryear{Frenk et al.}{1999}]{frenk99}
Frenk, C.S., White, S.D.M., Bode, P., et al., \aspj{1999}{525}{554}	

\bibitem[\protect\citeauthoryear{Fukugita et al.}{1998}]{fukugita98}
Fukugita, M., Hogan, C.J., \& Peebles, P.J.E., \aspj{1998}{503}{518}

\bibitem[\protect\citeauthoryear{Fusco-Femiano et al.}{2007}]{fusco-femiano07} 
Fusco-Femiano, R., Landi, R., \& Orlandini, M., \aspj{2007}{654}{L9}	

\bibitem[\protect\citeauthoryear{Gerhard et al.}{2005}]{gerhard05}
Gerhard, O., Arnaboldi, M., Freeman, K.C., et al., \aspj{2005}{621}{L93}	

\bibitem[\protect\citeauthoryear{Girardi et al.}{1998}]{girardi98}
Girardi, M., Giuricin, G., Mardirossian, F., Mezzetti, M., \& Boschin, W.,
\aspj{1998}{505}{74}

\bibitem[\protect\citeauthoryear{Gitti \& Schindler}{2004}]{gitti04}
Gitti, M., \& Schindler, S., \aeta{2004}{427}{L9}	

\bibitem[\protect\citeauthoryear{Gursky et al.}{1972}]{gursky72}
Gursky, H., Levinson, R., Kellogg, E., et al.,  \aspj{1972}{173}{L99}

\bibitem[\protect\citeauthoryear{Hanany et al.}{2000}]{hanany00}
Hanany, S., Ade, P., Balbi, A., et al., \aspj{2000}{545}{L5} 

\bibitem[\protect\citeauthoryear{Hayakawa et al.}{2006}]{hayakawa06}
Hayakawa, A., Hoshino, A., Ishida, M., et al., 2006,
PASJ,  58, 695	

\bibitem[\protect\citeauthoryear{Hennawi et al.}{2006}]{hennawi06}
Hennawi, J.F., Gladders, M.D., Oguri, M., et al., 2006, astro-ph/0610061	

\bibitem[\protect\citeauthoryear{Henry et al.}{2004}]{henry04}
Henry, J.P., Finoguenov, A., \& Briel, U.G., \aspj{2004}{615}{181}

\bibitem[\protect\citeauthoryear{Hernquist \& Katz}{1989}]{hernquist89}
Hernquist, L., \& Katz, N., \aspjs{1989}{70}{419}

\bibitem[\protect\citeauthoryear{Holmberg}{1941}]{holmberg41}
Holmberg, E., \aspj{1941}{94}{385}	

\bibitem[\protect\citeauthoryear{Jeltema et al.}{2005}]{jeltema05}
Jeltema, T.E., Canizares, C.R., Bautz, M.W., \& Buote, D.A., \aspj{2005}{624}{606}
 
\bibitem[\protect\citeauthoryear{Kaastra et al.}{2006}]{kaastra06}
Kaastra, J.S., Werner, N., den Herder, J.W.A., et al., 
\aspj{2006}{652}{189}	

\bibitem[\protect\citeauthoryear{Kaastra et al.}{2008}]{kaastra08}
Kaastra, J.S., Paerels, F.B.S., Durret, F., Schindler, S., \& Richter, P., 2008, SSR, in press

\bibitem[\protect\citeauthoryear{Kaiser}{1986}]{kaiser86}
Kaiser, N.,  \mnras{1986}{222}{323}

\bibitem[\protect\citeauthoryear{Kang et al.}{1994}]{kang94}
Kang, H., Ostriker, J.P., Cen, R., et al., \aspj{1994}{430}{83}	

\bibitem[\protect\citeauthoryear{Katz \& White}{1993}]{katz93}
Katz, N., \& White, S.D.M., \aspj{1993}{412}{455}	

\bibitem[\protect\citeauthoryear{Kauffmann et al.}{1999}]{kauffmann99}
Kauffmann, G., Colberg, J.M., Diaferio, A., \& White, S.D.M., \mnras{1999}{303}{188}	

\bibitem[\protect\citeauthoryear{Kodama et al.}{2006}]{kodama06}
Kodama, T., Tanaka, M., Kajisawa, M., et al., 2006, in 
Galaxy evolution across the Hubble time, IAU
Symp. 235, eds. F. Combes \& J. Palous, p. 27

\bibitem[\protect\citeauthoryear{Lee \& Evrard}{2007}]{lee07}
Lee, J., \& Evrard, A. E., \aspj{2007}{657}{30}	

\bibitem[\protect\citeauthoryear{Lieu \& Mittaz}{2005}]{lieu05}
Lieu, R., \& Mittaz, J.P.D., 2005, The Cluster soft excess: new faces of an old enigma, in 
 The identification of dark matter, eds. N.J.C. Spooner \& 
 V. Kudryavtsev, World Scient. Pub. Comp., Singapore, p. 18

\bibitem[\protect\citeauthoryear{Lynds \& Petrosian}{1986}]{lynds86}
Lynds, R., \& Petrosian, V., 1986,  BAAS,  18, 1014

\bibitem[\protect\citeauthoryear{Markevitch et al.}{2002}]{markevitch02}
Markevitch, M., Gonzalez, A.H., David, L., et al.,
\aspj{2002}{567}{L27}

\bibitem[\protect\citeauthoryear{McCarthy et al.}{2007}]{mccarthy07}
McCarthy, I.G., Bower, R.G., Balogh, M.L., et al., \mnras{2007}{376}{497}  

\bibitem[\protect\citeauthoryear{McNamara \& Nulsen}{2007}]{mcnamara07}
McNamara, B.R., \& Nulsen, P.E.J., 2007, ARA\&A, 45, 117 

\bibitem[\protect\citeauthoryear{McNamara et al.}{2006}]{mcnamara06}
McNamara, B.R., Rafferty, D.A., B\^\i rzan, L., et al., \aspj{2006}{648}{164}	

\bibitem[\protect\citeauthoryear{Metcalfe et al.}{2005}]{metcalfe05}
Metcalfe, L., Fadda, D., \& Biviano, A., \SpaceS{2005}{119}{425}

\bibitem[\protect\citeauthoryear{Metzler \& Evrard}{1994}]{metzler94}
Metzler, C.A., \& Evrard, A.E., \aspj{1994}{437}{564}	

\bibitem[\protect\citeauthoryear{Muchovej et al.}{2007}]{muchovej07}
Muchovej, S., Mroczkowski, T., Carlstrom, J.E., et al., \aspj{2007}{663}{708} 

\bibitem[\protect\citeauthoryear{Murante et al.}{2007}]{murante07}
Murante, G., Giovalli, M., Gerhard, O., et al., \mnras{2007}{377}{2}	

\bibitem[\protect\citeauthoryear{Navarro et al.}{1995a}]{navarro95a}
Navarro, J.F., Frenk, C.S., \& White, S.D.M., \mnras{1995a}{275}{56}

\bibitem[\protect\citeauthoryear{Navarro et al.}{1995b}]{navarro95b}
Navarro, J.F., Frenk, C.S., \& White, S.D.M., \mnras{1995b}{275}{720}

\bibitem[\protect\citeauthoryear{Nevalainen et al.}{2007}]{nevalainen07}
Nevalainen, J., Bonamente, M., \& Kaastra, J.,  \aspj{2007}{656}{733}	

\bibitem[\protect\citeauthoryear{Nicastro et al.}{2005}]{nicastro05}
Nicastro, F., Mathur, S., Elvis, M., et al., 2005,  Nature, 433, 495	

\bibitem[\protect\citeauthoryear{O'Shea et al.}{2005}]{oshea05}
O'Shea, B.W., Nagamine, K., Springel, V., Hernquist, L., \& Norman, M.L.,
\aspjs{2005}{160}{1}	

\bibitem[\protect\citeauthoryear{Paerels et al.}{2008}]{paerels08}
Paerels, F., Kaastra, J.S., Ohashi, T., et al., 2008, SSR, in press

\bibitem[\protect\citeauthoryear{Peebles}{1970}]{peebles70}
Peebles, P.J.E., \aj{1970}{75}{13}	

\bibitem[\protect\citeauthoryear{Petrosian et al.}{2006}]{petrosian06}
Petrosian, V., Madejski, G., \& Luli, K., \aspj{2006}{652}{948}	

\bibitem[\protect\citeauthoryear{Petrosian \& Bykov}{2008}]{petrosian08b}
Petrosian, V., \& Bykov, A.M., 2008, SSR, in press 

\bibitem[\protect\citeauthoryear{Petrosian et al.}{2008}]{petrosian08a}
Petrosian, V., Rephaeli, Y., \& Bykov, A.M., 2008, SSR, in press 

\bibitem[\protect\citeauthoryear{Plionis \& Basilakos}{2002}]{plionis02}
Plionis, M., \& Basilakos, S., \mnras{2002}{329}{L47}

\bibitem[\protect\citeauthoryear{Plionis et al.}{2003}]{plionis03}
Plionis, M., Benoist, C., Maurogordato, S., Ferrari, C., \& Basilakos, S., \aspj{2003}{594}{144}

\bibitem[\protect\citeauthoryear{Postman et al.}{2005}]{postman05}
Postman, M., Franx, M., Cross, N.J.G., et al., \aspj{2005}{623}{721}	

\bibitem[\protect\citeauthoryear{Press \& Schechter}{1974}]{press74}
Press, W.H., \& Schechter, P., \aspj{1974}{187}{425}	

\bibitem[\protect\citeauthoryear{Ramella et al.}{2007}]{ramella07}
Ramella, M., Biviano, A., Pisani, A., et al., \aeta{2007}{470}{39} 

\bibitem[\protect\citeauthoryear{Rauch et al.}{1997}]{rauch97}
Rauch, M., Miralda-Escud\'e, J., Sargent, W.L.W., et al., \aspj{1997}{489}{7}

\bibitem[\protect\citeauthoryear{Reiprich \& B\"ohringer}{2002}]{reiprich02}
Reiprich, T.H., \& B\"ohringer, H., \aspj{2002}{567}{716}	

\bibitem[\protect\citeauthoryear{Rephaeli et al.}{2008}]{rephaeli08}
Rephaeli, Y., Nevalainen, J., Ohashi, T., \& Bykov, A., 2008, SSR, in press

\bibitem[\protect\citeauthoryear{Richter et al.}{2008}]{richter08}
Richter, P., Paerels, F., \& Kaastra, J.S., 2008, SSR, in press

\bibitem[\protect\citeauthoryear{Rines et al.}{2007}]{rines07}
Rines, K., Finn, R., \& Vikhlinin, A., \aspj{2007}{665}{L9} 

\bibitem[\protect\citeauthoryear{Rines et al.}{2005}]{rines05}
Rines, K., Geller, M.J., Kurtz, M.J., \& Diaferio, A., \aj{2005}{130}{1482}	

\bibitem[\protect\citeauthoryear{Rines et al.}{2001}]{rines01}
Rines, K., Mahdavi, A., Geller, M.J., et al., \aspj{2001}{555}{558}

\bibitem[\protect\citeauthoryear{Rosati}{2004}]{rosati04} 
Rosati, P., 2004, in Clusters of galaxies: probes of cosmological structure and galaxy evolution, 
Cambridge Univ. Press, Carnegie Obs. 
Astrophys. Ser., eds. J.S. Mulchaey, A. Dressler, \& A. Oemler, p. 72

\bibitem[\protect\citeauthoryear{Rosati et al.}{2002}]{rosati02}
Rosati, P., Borgani, S., \& Norman, C., \araa{2002}{40}{539}

\bibitem[\protect\citeauthoryear{Rossetti \& Molendi}{2004}]{rossetti04}
Rossetti, M., \& Molendi, S., \aeta{2004}{414}{L41}

\bibitem[\protect\citeauthoryear{Roukema et al.}{1997}]{roukema97}
Roukema, B.F., Quinn, P.J., Peterson, B.A., \& Rocca-Volmerange, B., \mnras{1997}{292}{835}	

\bibitem[\protect\citeauthoryear{Sanders et al.}{2005}]{sanders05}
Sanders, J.S., Fabian, A.C., \& Dunn, R.J.H., \mnras{2005}{360}{133}

\bibitem[\protect\citeauthoryear{Sastry}{1968}]{sastry68}
Sastry, D.N., 1968, PASP,  80, 252

\bibitem[\protect\citeauthoryear{Schindler \& Diaferio}{2008}]{schindler08}
Schindler, S., \& Diaferio, A., 2008, SSR, in press

\bibitem[\protect\citeauthoryear{Schneider}{2006}]{schneider06}
Schneider, P., 2006, in Gravitational lensing: strong, weak and micro. Saas-Fee Advanced Course 33, 
eds. G. Meylan, P. Jetzer \& P. North, Springer, Berlin, p. 1

\bibitem[\protect\citeauthoryear{Schuecker et al.}{2001}]{schuecker01}
Schuecker, P., B\"ohringer, H., Reiprich, T. H., \& Feretti, L., \aeta{2001}{378}{408}	

\bibitem[\protect\citeauthoryear{Schuecker et al.}{2004}]{schuecker04}
Schuecker, P., Finoguenov, A., Miniati, F., B\"ohringer, H., \& Briel, U.G., \aeta{2004}{426}{387}	

\bibitem[\protect\citeauthoryear{Shapley}{1926}]{shapley26}
Shapley, H., 1926, Harvard. Obs. Bull.,  838, 3

\bibitem[\protect\citeauthoryear{Shapley}{1933}]{shapley33}
Shapley, H., 1933, Proc. Nat. Acad. Sci. Washington,  19, 591

\bibitem[\protect\citeauthoryear{Sheth \& Diaferio}{2001}]{sheth01}
Sheth, R.K., \& Diaferio, A., \mnras{2001}{322}{901}

\bibitem[\protect\citeauthoryear{Sheth \& Tormen}{1999}]{sheth99}
Sheth, R.K., \& Tormen, G., \mnras{1999}{308}{119}	

\bibitem[\protect\citeauthoryear{Smith et al.}{2005}]{smith05}
Smith, G.P., Treu, T., Ellis, R.S., Moran, S.M., \& Dressler, A., \aspj{2005}{620}{78}	

\bibitem[\protect\citeauthoryear{Smoot et al.}{1992}]{smoot92}
Smoot, G.F., Bennet, C.L., Kogut, A., et al., \aspj{1992}{396}{L1}

\bibitem[\protect\citeauthoryear{Spergel et al.}{2007}]{spergel07}
Spergel, D.N., Bean, R., Dor\'e, O., et al., \aspjs{2007}{170}{377}

\bibitem[\protect\citeauthoryear{Springel et al.}{2005}]{springel05}
Springel, V., White, S.D.M., Jenkins, A., et al., 2005,  Nature,  435, 629	

\bibitem[\protect\citeauthoryear{Springel et al.}{2006}]{springel06}
Springel, V., Frenk, C.S., \& White, S.D.M., 2006, Nature, 440, 1137

\bibitem[\protect\citeauthoryear{Sunyaev \& Zeldovich}{1972}]{sunyaev72}
Sunyaev, R.A., \& Zeldovich, Ya.B., 1972, Comm. Astroph. Sp. Phys., 4, 173

\bibitem[\protect\citeauthoryear{Takei et al.}{2007}]{takei07}
Takei, Y., Henry, J.P., Finoguenov, A., et al., \aspj{2007}{655}{831}

\bibitem[\protect\citeauthoryear{Tegmark et al.}{2004}]{tegmark04}
Tegmark, M., Blanton, M.R., Strauss, M.A., et al., \aspj{2004}{606}{702}

\bibitem[\protect\citeauthoryear{The \& White}{1986}]{the86}
The, L.S., \& White, S.D.M., \aj{1986}{92}{1248}

\bibitem[\protect\citeauthoryear{Thomas \& Couchman}{1992}]{thomas92}
Thomas, P.A., \& Couchman, H.M.P., \mnras{1992}{257}{11}	

\bibitem[\protect\citeauthoryear{Valdarnini}{2002}]{valda02}
Valdarnini, R., \aspj{2002}{567}{741}	

\bibitem[\protect\citeauthoryear{Valdarnini}{2003}]{valda03}
Valdarnini, R., \mnras{2003}{339}{1117}	

\bibitem[\protect\citeauthoryear{Voit}{2005}]{voit05}
Voit, G.M., \rvmp{2005}{77}{207}

\bibitem[\protect\citeauthoryear{von Hoerner}{1960}]{hoerner60}	
von Hoerner, S., 1960,  Z. Astrophys.,  50, 184

\bibitem[\protect\citeauthoryear{Werner et al.}{2008}]{werner08}
Werner, N., Durret, F., Ohashi, T., Schindler, S., \& Wiersma, R.P.C., 2008, SSR, in press

\bibitem[\protect\citeauthoryear{Werner et al.}{2007}]{werner07}
Werner, N., Kaastra, J.S., Takei, Y., et al., \aeta{2007}{468}{849} 

\bibitem[\protect\citeauthoryear{West}{1994}]{west94}
West, M.J., \mnras{1994}{268}{79}	

\bibitem[\protect\citeauthoryear{White}{1976}]{white76}
White, S.D.M., \mnras{1976}{177}{717}	

\bibitem[\protect\citeauthoryear{Wolf}{1902}]{wolf02}
Wolf, M., 1902, Publik. des Astrophys. Inst. K\"onigstuhl-Heidelberg,  1, 125

\bibitem[\protect\citeauthoryear{York et al.}{2000}]{york00}
York, D.G., Adelman, J., Anderson, J.E., Jr., et al., \aj{2000}{120}{1579}

\bibitem[\protect\citeauthoryear{Zibetti et al.}{2005}]{zibetti05}
Zibetti, S., White, S.D.M., Schneider, D.P., \& Brinkmann, J., \mnras{2005}{358}{949}	

\bibitem[\protect\citeauthoryear{Zwicky}{1933}]{zwicky33}
Zwicky, F.,  1933,  Helvetica Physica Acta,  6, 110

\bibitem[\protect\citeauthoryear{Zwicky}{1937}]{zwicky37}
Zwicky, F., \aspj{1937}{86}{217}

\bibitem[\protect\citeauthoryear{Zwicky et al.}{1961-68}]{zwicky68}
Zwicky, F., Herzog, E., Wild, P., Karpowicz, M., \& Kowal C.T., 1961-68,  Catalogue of
Galaxies and Clusters of Galaxies, Pasadena, California Institute of Technology

\end{thebibliography}
\end{document}